\documentclass[%
 reprint,
 amsmath,
 amssymb,
 aps,
 prmaterials,
 longbibliography,
]{revtex4-2}

\usepackage{graphicx}
\usepackage{dcolumn}
\usepackage{bm}
\usepackage{siunitx}
\usepackage{prettyref}

\DeclareSIUnit\angstrom{\text {\AA}}

\newrefformat{tab}{Table \ref{#1}}
\newrefformat{fig}{Figure \ref{#1}}

\DeclareUnicodeCharacter{2212}{-}

\begin{document}

\title{Dynamically training machine-learning-based force fields for strongly anharmonic materials}

\author{Martin Callsen$^1$}
 \email{mcallsen@as.edu.tw}
\author{Tai-Ting Lee$^{1,2}$}
\author{Mei-Yin Chou$^{1,2}$}%
 \email{mychou6@as.edu.tw}
\affiliation{%
 $^1$Institute of Atomic and Molecular Sciences, Academia Sinica, Taipei 10617, Taiwan\\
 $^2$Department of Physics, National Taiwan University, Taipei 10617, Taiwan
}

\date{\today}

\begin{abstract}
Machine learning (ML) force fields have emerged as a powerful tool for computing materials properties at finite temperatures, particularly in regimes where traditional phonon-based perturbation theories fail or cannot be extended beyond the harmonic approximation. These approaches offer accuracy comparable to ab initio molecular dynamics (MD), but at a fraction of the computational cost. However, their reliability critically depends on the quality and representativeness of the training data. In particular, static training datasets often lead to failure when the force field encounters previously unseen atomic configurations during MD simulations.
In this work, we present a framework for dynamically training ML force fields and demonstrate its effectiveness across materials with varying degrees of anharmonicity, including cubic boron arsenide (c-BAs), silicon (Si), and tin selenide (SnSe). Our method builds on the conventional lattice dynamics expansion of total energy and incorporates Bayesian error estimation to guide adaptive data acquisition during simulation. Specifically, we show that trajectory-averaged Bayesian errors enable efficient and targeted exploration of the configuration space, significantly enhancing the robustness and transferability of the resulting force fields. We further demonstrate how Bayesian error estimation can be applied to determine the convergence of the dynamic training without requiring additional ab initio data. This proposed framework offers a practical and easily implementable scheme to improve the training process, which is the most critical step in developing reliable ML force fields.
\end{abstract}


\maketitle

\section{\label{sec:introduction} Introduction}

Including anharmonic effects has become crucial when computing materials properties at finite temperatures that would otherwise diverge or vanish in the harmonic approximation, such as thermal conductivity or thermal expansion. In particular, ultra-low lattice thermal conductivity, which is relevant for technological applications as thermoelectrics, tends to occur in strongly anharmonic materials. This can be attributed to their complicated potential energy surface (PES) featuring energetically similar structures or alternate structural phases. At the same time, computing materials properties at finite temperatures for anharmonic materials is challenging, since in this case commonly used phonon-based perturbation theories break down \cite{PhysRevMaterials.4.083809, PhysRevB.110.235202} and going beyond the harmonic or quasi-harmonic approximation is, while in principle possible, costly and tedious. As an alternative approach, in principle the properties mentioned above can be obtained from sufficiently long \textit{ab initio} molecular dynamics (AIMD) trajectories \cite{C2CP42394D, PhysRevB.94.054304, PhysRevLett.112.058501, PhysRevB.110.235202, PhysRevB.107.224304, PhysRevMaterials.9.045403}, since they encode the full information about the dynamics of the system. However, the computational cost of obtaining the said trajectories is in most cases prohibitively high. 

Recently, due to their promise of AIMD level accuracy at significantly reduced cost, machine-learning-based force fields trained on \textit{ab initio} or higher level forces have made obtaining the required trajectories feasible. In particular, machine learning interatomic potentials (MLIPs) \cite{AdvMater.31.1902765.2019, WANG2024109673}, which are based on neural networks, have gained substantial popularity due to their universal applicability. While there is a growing number of applications showcasing the method's potential such as relaxation of Moir\'e structures \cite{npjCompMaterials.10.169.2024, PhysRevMaterials.9.014004} and thermal properties \cite{PhysRevMaterials.9.045403}, in most of these cases further optimization, either in terms of specialized training data or empirical parameters, was required in order to achieve the desired accuracy. The main issue is that MLIPs are trained on extensive but static structural databases. While these databases already contain hundreds of thousands of structures, those are typically still restricted to the equilibrium structures and their close vicinity. Once the MLIPs are applied far away from their training domain, this can result in a general underestimation of the potential energy surface \cite{Deng2025}, an erroneous presence or absence of metastable states on the PES \cite{PhysRevMaterials.9.063801}, or, in the worst case, failure of the MD. This is a particular problem for anharmonic materials, where structures that significantly differ from the equilibrium structure will be encountered on a regular basis due to the aforementioned complexity of the PES.

The solution adopted by more and more MLIPs is to implement an active or dynamic training method \cite{PhysRevLett.114.096405, PhysRevLett.122.225701, PhysRevB.100.014105, npjCompMater.1.20.2020, PhysRevMaterials.9.063801} that searches the configuration space (CS) for unknown structures by MD and updates the training data accordingly. Such a scheme requires a measure of the current uncertainty of the predictions. However, depending on the architecture, most neural networks do not have access to a native uncertainty measure. Thus, a common approach is utilizing a so-called ensemble average \cite{PhysRevMaterials.9.063801}, which takes the deviation from the average over predictions made by models either trained on disjoint sets of training data or repeatedly trained on the same training data to assess the model's uncertainty. For force fields relying on linear regression \cite{PhysRevB.100.014105, 10.1063/5.0129045} instead of neural networks, Bayesian error estimation \cite{PhysRevB.100.014105, CBishop2006} provides a powerful way to quantify the uncertainty that combines complete versions of both the aforementioned training data and parameter ensembles. Given a suitable uncertainty measure, the next two pressing questions are when to retrain the force field in order to avoid redundant training data and most importantly when the dynamic training is sufficiently converged.

In this work, we present a dynamic force field training method combining an ML-force field obtained from the usual lattice dynamics expansion of the total energy and Bayesian error estimation as uncertainty measure, which will be detailed in Section~\ref{sec:method}. In Section~\ref{sec:results} we will first showcase an efficient way for selecting training data and reducing the overall uncertainty based on trajectory averages of the uncertainty measure, which will be applied to force fields for materials with varying degrees of anharmonicity, i.e. in increasing order BAs, Si and SnSe. Then we will discuss how Bayesian error estimation also enables us to simultaneously determine the convergence of the dynamic training without relying on comparison with \textit{ab initio} forces. Finally, we will assess the distribution of uncertainties in the accessible part of the CS in order to verify the previous observations.

\section{Methodology}
\subsection{\label{sec:method} Dynamic training method}

\begin{figure}[tb]
\includegraphics[width=\columnwidth]{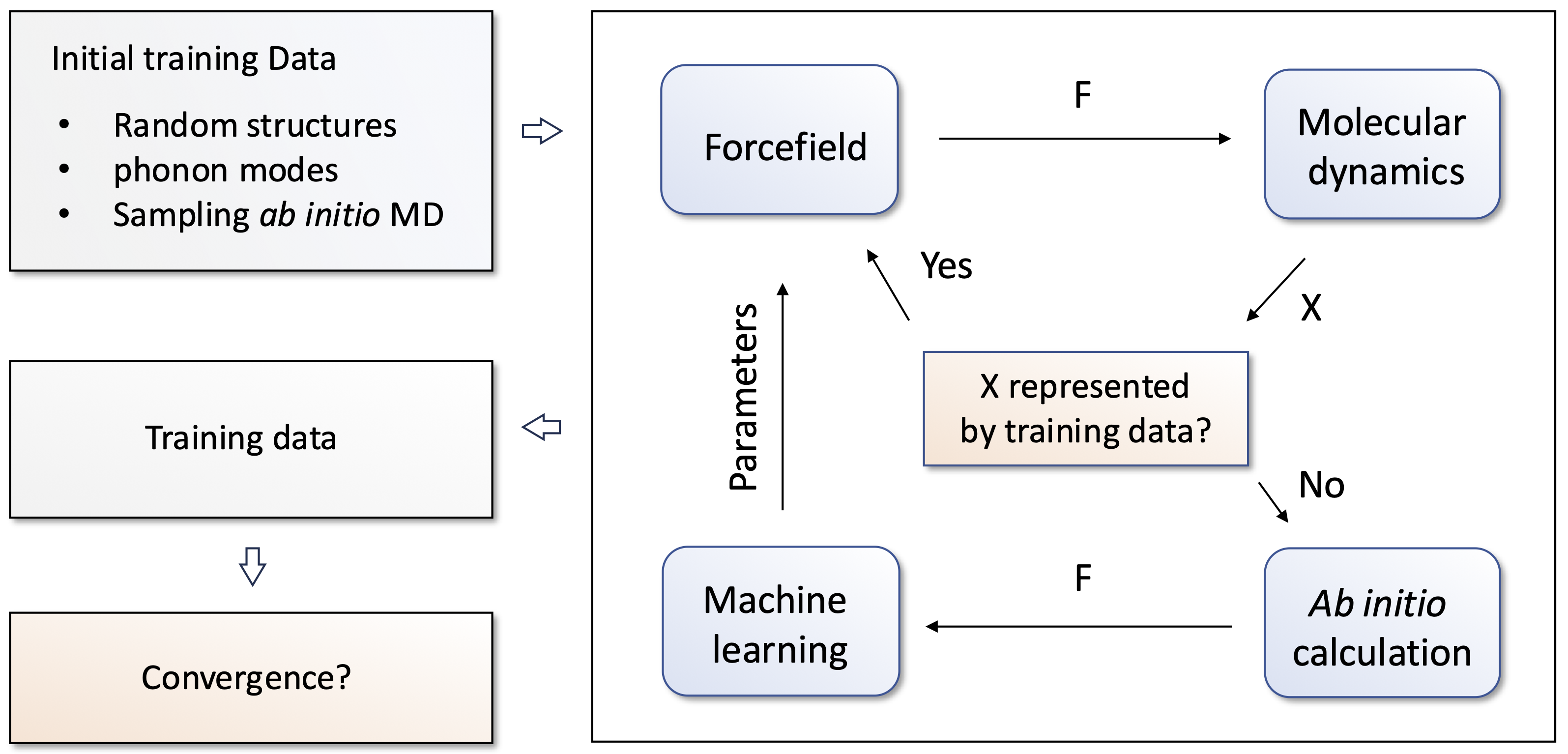}
\caption{
    \label{fig:figure_1}The flow-chart for the dynamic force field training method. The idea is to explore the configuration space by using molecular dynamics with a force field trained on some initial training data. Whenever a structure $\mathbf{X}$ that is not well-represented by the training data is encountered during the training, the force field will be updated based on \textit{ab initio} forces. 
}
\end{figure}

In the following, we will describe the general flow of the dynamic force field training method as outlined in Figure~\ref{fig:figure_1}. The main idea of this iterative scheme is exploring the PES by MD, in order to identify structures with high uncertainty in the predictions of the current force field. Whenever such a structure is encountered, we obtain the \textit{ab initio} forces in order to update the training data and retrain the force field. The initial training data can be obtained from random structures, displacing atoms along the phonon modes of the system, or sampled from a short \textit{ab initio} MD trajectory. Once a sufficient number of new training structures has been added, we also have to determine whether the dynamic training has converged, i.e., the force field has reached the desired quality.

The first component required for this iterative scheme is the force field itself, for which we employ compressive sensing lattice dynamics (CSLD) \cite{PhysRevLett.113.185501, PhysRevB.100.184308}. In this approach the potential energy $E$ of the system is given by the usual lattice-dynamics expansion in terms of the displacements $u_{i}^a$ of atom $a$ in cartesian direction $i$:
\begin{equation}
    \label{eq:ld_expansion}
    E = E_0 + \frac{1}{2!} \Phi_{ij}^{ab} u_{i}^{a} u_{j}^{b} + \frac{1}{3!} \Phi_{ijk}^{abc} u_{i}^{a} u_{j}^{b} u_{k}^{c} + \dots,
\end{equation}
where the expansion coefficients $\Phi$ are the $n$-th order force constant tensors (FCTs) of the system and summation over repeated indices is implied. The first-order term in this expansion vanishes in the equilibrium. The second-order expansion coefficients $\Phi_{ij}^{ab} = \partial^2 E / \partial u_i^a\partial u_j^b$ determine the phonon dispersion in the harmonic approximation. The third- and higher-order terms are the anharmonic FCTs, which are essential for thermal transport and finite phonon lifetimes. This expansion is in principle exact and provides us with a natural way to account for anharmonic effects up to arbitrary order $n$. In practice, however, we have to truncate the summation at a certain order, due to the exponentially growing number of FCTs. In addition, to keep the computations feasible, we will only include terms in the summation in Equation~\ref{eq:ld_expansion}, for which the size of the cluster of atoms $\mathbf{c} = \left\{ a, b, c, \ldots \right\}$ corresponding to the FCT $\Phi_{ijk\ldots}^{abc\ldots}$ is smaller than a certain cut-off radius $r_{\text{cut}}$. 

We obtain the force $\mathbf f^a = -\nabla_{\mathbf u^a}E$ acting on atom $a$ by taking the gradient of the total energy. While the expansion in Equation~\ref{eq:ld_expansion} is non-linear in the displacements of the atoms $\mathbf{u}$, it is linear in terms of the FCTs $\Phi$. Thus, by collecting all the components of the FCTs corresponding to each summand in Equation~\ref{eq:ld_expansion} in a vector $\boldsymbol{\phi}$ we can recast the equation for the forces $\mathbf{f} \equiv \{ \mathbf{f}^a,  \mathbf{f}^b, \ldots\}$ acting on the atoms in a form suitable for linear regression \cite{PhysRevB.100.184308}: 
\begin{equation}
\label{eq:forces}
    \mathbf{f} = \mathrm{A} (\mathbf{u}) \, \boldsymbol{\phi} = \mathrm{A} (\mathbf{u}) \, \mathrm{C} \, \boldsymbol{\phi}^\prime.
\end{equation}
The elements of the matrix $\mathrm A$ in this expression contain the products of the displacements $u_{i}^{a} u_{j}^{b} u_{k}^{c} \ldots$ for the corresponding component $\Phi_{ijk\ldots}^{abc\ldots}$ in $\boldsymbol{\phi}$. Many of the components of $\boldsymbol{\phi}$ are equivalent by one of the systems symmetries, i.e. permutation symmetry, space-group symmetries, or translational invariance, which are encoded in the $\mathrm C$ mapping between the dependent parameters $\boldsymbol{\phi}$ and the independent parameters $\boldsymbol{\phi}^\prime$. Even for systems with low space-group symmetry, the number of independent FCTs can be as much as an order of magnitude smaller than the total number of FCTs \cite{PhysRevB.100.184308}. The components of the FCTs $\boldsymbol{\phi}$ can then be directly obtained as the parameters of a linear regression model fitted to \textit{ab initio} forces. Thus, we can directly interpret and extract information from the parameters of our model, which is one of the major advantages over choosing a neural network based MLIP.

The second component required for the dynamic training scheme is an uncertainty measure in order to determine whether or not the forces $\mathbf f$ at any given MD step are well-represented by the training data consisting of \textit{ab initio} forces $\mathbf{F}$. Choosing a force field that is based on linear regression allows us to employ Bayesian error estimation for this purpose, as described in detail in \cite{PhysRevB.100.014105, PhysRevLett.122.225701}. For the purposes of this manuscript, within the Bayesian formulation of linear regression the probability of predicting forces $\mathbf{f}$ for displacements $\mathbf{u}$ from a model based on training data $\mathbf{F}$ is given by the \textit{predictive} distribution $p(\mathbf{f} \, | \, \mathbf{F}, \alpha, \beta)$, where $\alpha$ and $\beta$ are hyper-parameters describing the width of the distribution of parameters $\boldsymbol{\phi}$ and the deviation of the predictions with respect to the training data, respectively. Under the assumption that both of these \textit{prior} distributions are Gaussian, also the \textit{predictive} distribution will be a normal distribution 
\begin{equation}
    \label{eq:probability}
    p(\mathbf{f} \, | \, \mathbf{F}, \alpha, \beta) = \mathcal{N} (\mathbf{f} \, | \, \mathrm{A} (\mathbf{u}) \, \overline{\boldsymbol{\phi}}, \sigma^2),
\end{equation}
which is centered around the forces $\overline{\mathbf{f}} = \mathrm{A} (\mathbf{u}) \, \overline{\boldsymbol{\phi}}$ that the model would predict with the optimized parameters $\overline{\boldsymbol{\phi}}$. The variance $\sigma^2$ of this distribution is
\begin{equation}
    \label{eq:sigma}
    \sigma^2 = \beta^{-1} \, \mathbb{I} + \mathrm{A} (\mathbf{u})  \, \Sigma (\mathbf{F}) \, \mathrm{A}^\intercal (\mathbf{u}),
\end{equation}
where $\Sigma (\mathbf{F})$ is the matrix representing the width of the distribution of parameters that could be obtained with the current training data.

The first term of $\sigma^2$ measuring the deviation of the predictions from the training data is constant over the entire CS and only changes when the training data is updated. The second term in contrast, the square root of which will be denoted as Bayesian error $\varepsilon_{\text B}$ in the following, depends on the current MD step through $\mathrm{A} (\mathbf{u})$ and, as has been shown previously, closely follows the actual error of the forces \cite{PhysRevB.100.014105}. This can be understood by considering that a large $\varepsilon_{\text B}$ corresponds to a high probability of obtaining very different forces  when changing the parameters $\boldsymbol{\phi}$. In other words, this means that the predicted forces are very likely to be wrong. 
The major advantage of this approach as already pointed out in \cite{PhysRevB.100.014105} is that the Bayesian error provides a good estimation of the actual error of the forces without ever explicitly calculating the corresponding \textit{ab initio} forces. In particular, this means that the training does not require an actual \textit{ab initio} MD trajectory. Instead, the trajectory from which the training data is selected can be efficiently generated with the current force field, allowing for significantly longer training trajectories. 

As a final remark about the Bayesian error estimation we would like to note that the above assumption of a Gaussian \textit{prior} distribution for the parameters corresponds to Ridge regression i.e., using the L$_2$ norm as a regularizer. For other choices of the distribution of the parameters, e.g. a Laplacian distribution for the case of LASSO regression, which was used in the original implementation of CSLD \cite{PhysRevLett.113.185501, PhysRevB.100.184308}, the integrals in the derivation of Equation~\ref{eq:probability} become intractable. Thus, in order to make use of Bayesian error estimation, our implementation of the force field relies on Ridge regression instead.

With a suitable uncertainty measure at hand, the next crucial step is deciding when to retrain the force field based on that in order to keep the dynamic training efficient and the amount of training data at a manageable level. In the original implementation \cite{PhysRevB.100.014105} the Bayesian error $\varepsilon_{\text{min}}$ evaluated at the MD step directly after retraining the force field is stored as a representation for the currently lowest achievable error and the force field is retrained whenever the current $\varepsilon_{\text B}$ is larger than the average of a number of stored $\varepsilon_{\text{min}}$. As will be shown in the next section, this is a somewhat strict criterion.  While it does lead to an overall decreasing $\varepsilon_{\text B}$ throughout the training, it will pick up a large number of redundant structures that only marginally improve the force field. Thus, we chose an alternative approach by instead comparing $\varepsilon_{\text B}$ at the current MD step to a trajectory average of the Bayesian error $\mu_k[\varepsilon_{\text B}]$ taken over the previous $k$ MD steps. The force field will then be updated whenever
\begin{equation}
    \label{eq:criterion}
    \varepsilon_{\text B} > \gamma \, \mu_k[\varepsilon_{\text B}].
\end{equation}
For the force fields trained in this work $\gamma = 3$ resulted in a reasonable amount of training data while keeping the dynamic training stable. 

The idea behind this trajectory average criterion is that the MD trajectory follows a landscape of regions with varying average uncertainty in the CS. The regions with the lowest uncertainty are centered around the training structures. The purpose of the dynamic training then is to identify structures with unusually high $\varepsilon_{\text B}$ within their respective region of the CS. Recently, an active learning scheme has been reported \cite{PhysRevMaterials.9.063801} that utilizes an acquisition function based on the average of their uncertainty measure. The average is taken over a test set sampled from an \textit{ab initio} MD trajectory and only updated prior to each round of active training. While this method is conceptually similar to the approach proposed in this work, keeping the average fixed during the MD means that the choice of the test set defines what is considered high or low uncertainty. The trajectory average employed in our method is designed to follow the aforementioned uncertainty landscape and thus provides a more adaptive way of selecting new training structures.

\subsection{Computational details}

\begin{table}
\caption{\label{tab:computational_details}Computational details for the force field expansion and the dynamic training. The anharmonicity measure $\sigma_{\text{A}}$ at $300~\text{K}$ as defined in \cite{PhysRevMaterials.4.083809}, maximum order $n_{\text{max}}$ and cut off radii $r_{\text{cut}}$ for the second and higher order terms in the cluster expansion, and the number structures in the training data $N_{\text{data}}$ after $500~\text{ps}$ for the force field training of BAs, Si, and SnSe.}
\begin{ruledtabular}
\begin{tabular}{lccc}
 & BAs & Si & SnSe\\
\hline
 $\sigma_{\text{A}}$ $300~\text{K}$ & $0.12$ & $0.14$ & $0.32$ \\
$n_{\text{max}}$ & $4$ & $4$ & $6$ \\
$r_{\text{cut}}^{2}$ (\AA) & $7.1$ & $8.0$ & $11.2$ \\
$r_{\text{cut}}^{3+4}$ (\AA) & $4.0$ & $5.0$ & $4.5$ \\
$r_{\text{cut}}^{5+6}$ (\AA) & & & $3.4$ \\
$N_{\text{data}}$ & $81$ & $60$ & $314$ \\
\end{tabular}
\end{ruledtabular}
\end{table}

The choice of the maximum order $n_{\text{max}}$ depends on the anharmonicity of the material, which can be quantified by the anharmonicity measure $\sigma_{\text{A}}$ as defined by Knoop \textit{et al.} \cite{PhysRevMaterials.4.083809}. For the example materials with varying degrees of anharmonicity considered in this work, specifically cubic BAs, Si, and SnSe, the respective $\sigma_{\text{A}}$ at $300~\text{K}$ are $0.12$, $0.14$, and $0.32$. Due to the considerably higher anharmonicity of SnSe, the corresponding force field has been expanded up to the sixth order, while for BAs and Si an expansion up to the fourth order was sufficient. As summarized in Table~\ref{tab:computational_details}, the cut-off radii for the second order terms were chosen as $7.1~\text{\AA}$, $8.0~\text{\AA}$, and $11.2~\text{\AA}$, respectively. For the third and fourth order terms the cut-off radii were $4.0~\text{\AA}$, $5.0~\text{\AA}$, and $4.5~\text{\AA}$ corresponding to third nearest neighbor interaction for BAs and Si and second nearest neighbor for SnSe. 

The MD is taken care of by the Atomic Simulation Environment (ASE) \cite{HjorthLarsen_2017}, which implements a variety of different thermostats for the NVT ensemble. During the training we use a Langevin thermostat to keep the system at the desired temperature. As an additional benefit, the non-deterministic nature of this particular thermostat ensures a faster exploration of the configuration space during training. The time step for the MD simulations was $1~\text{fs}$.

Once a new structure for the training has been identified, the corresponding \textit{ab initio} forces have to be calculated e.g. from density functional theory (DFT). In this particular case they have been obtained using the Vienna Ab Initio Simulation Package (VASP) \cite{Kresse:1996,Kresse:1999}. The electronic structure calculations are embedded within our computational framework by the ASE meaning that in principle any other suitable DFT package with a corresponding interface could be used. Since the training requires accurate forces, the computational parameters have to be chosen accordingly. In particular, the plane-wave cut-off energies for Si was $350~\text{eV}$ and $500~\text{eV}$ for BAs and SnSe, and as an approximation for the xc-functional the PBEsol functional \cite{PhysRevLett.100.136406} has been chosen, which has proven to yield reasonable structural parameters and lattice dynamics for a range of semiconductors and anharmonic materials \cite{JChemPhys.143.064710.2015, PhysRevLett.117.075502}. The supercell size for the training is determined by the choice of cut-off radii in the cluster expansion. For BAs and Si a $3 \times 3 \times 3$ supercell of the conventional cell with 216 atoms and for SnSe a $6 \times 8 \times 2$ supercell with 768 atoms have been chosen. Finally, the corresponding $\mathbf k$-point grids for the supercell calculations were $2 \times 2 \times 2$  Monkhorst-Pack grids for BAs and Si as well as a $2 \times 2 \times 4$ grid for SnSe.

\section{Results and Discussion}
\subsection{\label{sec:results} Training force fields for materials with varying anharmonicity}

\begin{figure}[tb]
\includegraphics[width=\columnwidth]{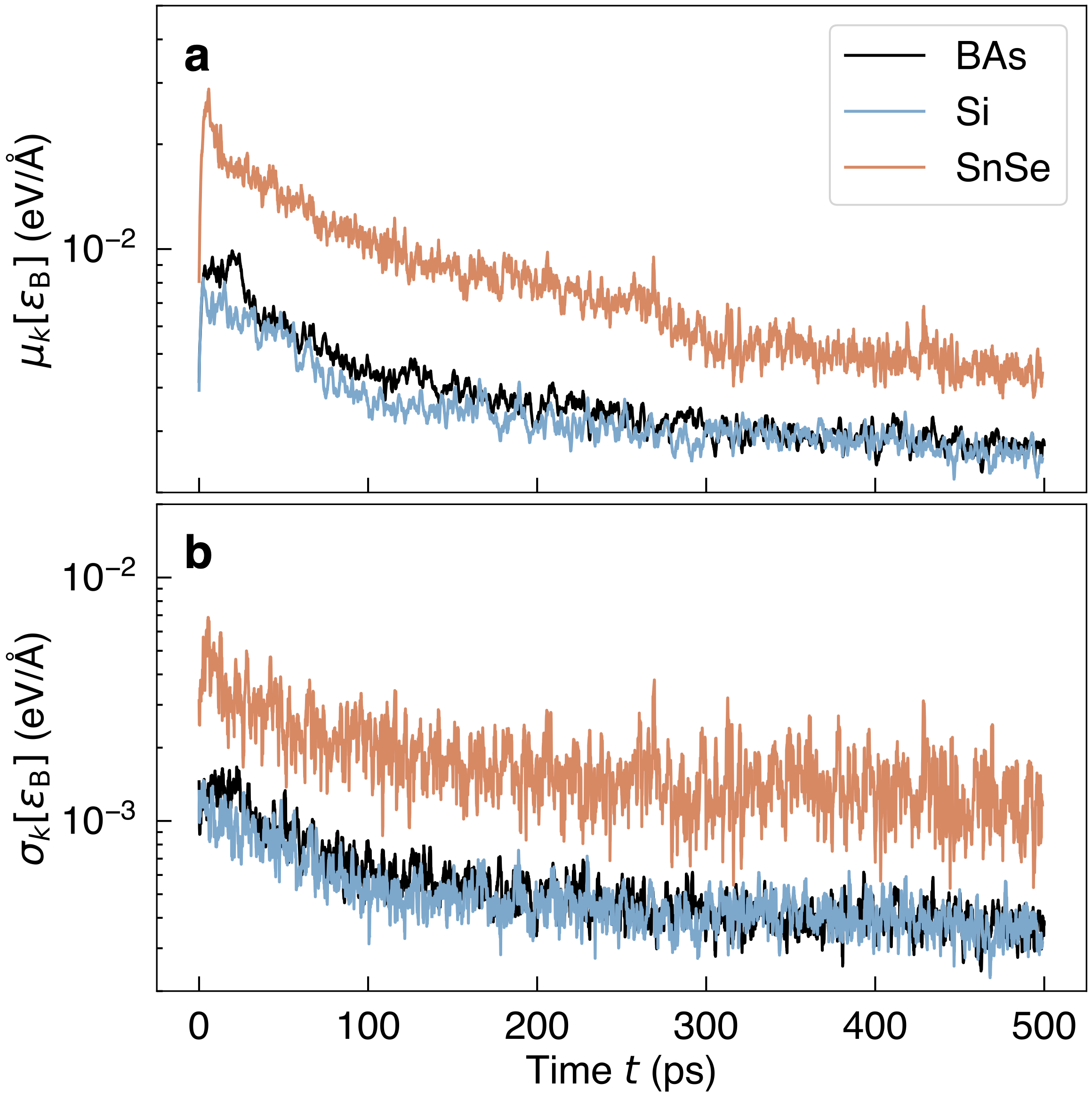}
\caption{
    \label{fig:figure_2}The trajectory average $\mu_k[\varepsilon_{\text B}]$ of the Bayesian error (a) and the corresponding standard deviation $\sigma_k[\varepsilon_{\text B}]$ during the dynamic training. Both quantities have been evaluated for the $500~\text{ps}$ training trajectories of BAs (black), Si (blue), and SnSe (red) and the averages have been taken over $k = 1000$ molecular dynamics steps.
}
\end{figure}  

In the following, we discuss the dynamic training of the force fields for BAs, Si, and SnSe using the trajectory average criterion in Equation~\ref{eq:criterion} as described in the previous section. This set of materials represent systems with increasing levels of anharmonicity. In Figure~\ref{fig:figure_2} (a), we show the trajectory average $\mu_k[\varepsilon_{\text B}]$ of the Bayesian error during the dynamic training
for these three materials. It can be seen that once the system has reached the desired temperature, the average Bayesian error assumes its maximum and from then on decreases rapidly. After $500~\text{ps}$ of dynamic training, the average Bayesian error for BAs was $\approx 3~\text{meV/\AA}$ and the final training data set contained $81$ structures. For comparison, a force field trained with the original $\varepsilon_{\text{min}}$ criterion \cite{PhysRevB.100.014105} ended up with $105$ structures when reaching the same level of Bayesian error. The corresponding standard deviation $\sigma_k[\varepsilon_{\text B}]$ shown in Figure~\ref{fig:figure_2} (b) amounts to approximately one third of $\mu_k[\varepsilon_{\text B}]$ for the case of BAs. In combination with our choice of $\gamma = 3 $ for the refit criterion in Equation~\ref{eq:criterion} this implies that newly added training structures are $\approx 9$ standard deviations away from the current average and thus have an unusually high uncertainty for their respective region of the CS. While Si behaves similar to BAs,  the dynamic training for strongly anharmonic SnSe has picked up 314 training structures even with our more efficient trajectory average criterion, and the Bayesian error at the end was $\approx 6~\text{meV/\AA}$. We would first of all like to note that even though compared to the more harmonic BAs and Si the error is significantly higher, the resulting force field is stable enough to run MD with the dynamic training turned off for more than $500~\text{ps}$. On the one hand, this is a significant improvement over the statically trained force fields mentioned in the introduction, which were failing after just a couple of pico seconds. On the other hand however, the large variance of the Bayesian error for materials with different anharmonicity makes the results of the dynamic training incomparable. In addition, $\varepsilon_{\text{B}}$ is supposed to monotonically decrease to zero for large numbers of training data \cite{CBishop2006}, which in combination means that we will not be able to determine the convergence of the dynamic training or the quality of the force field based on $\varepsilon_{\text{B}}$ alone. 

\subsection{Convergence of the dynamic training}

\begin{figure}[tb]
\includegraphics[width=\columnwidth]{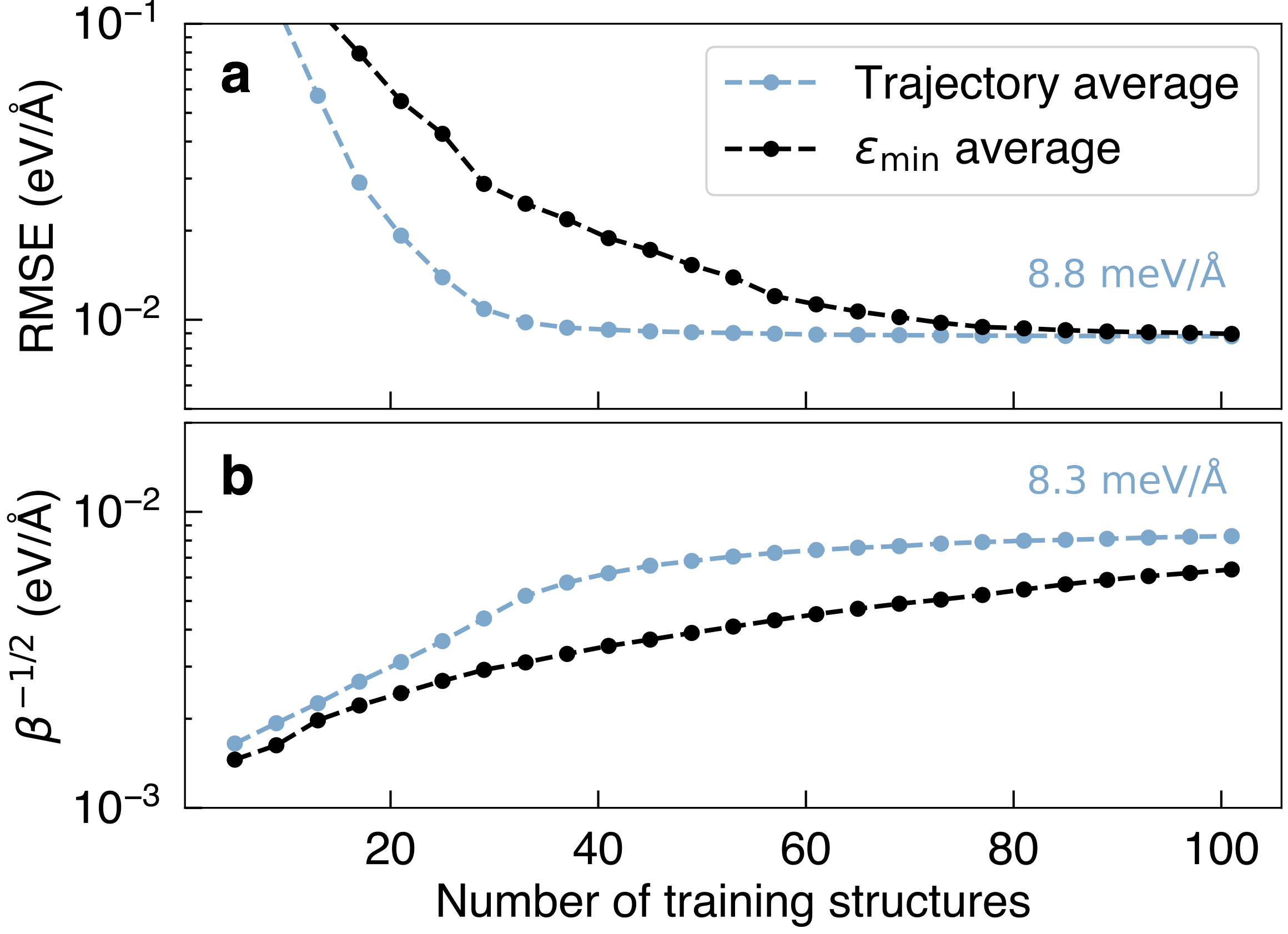}
\caption{
    \label{fig:figure_3}Convergence of the RMSE with respect to \textit{ab inito} forces on a test set (a) and the fitting error $\beta^{-1/2}$ of the linear regression (b) during the dynamic training of the BAs force-field as a function of the number of training structures. The force-fields have been trained using either the $\varepsilon_{\text{min}}$ (black) criterion or the trajectory average criterion (blue). The test set with \textit{ab initio} forces comprising 50 structures was sampled from a $150~\text{ps}$ NVT trajectory at $300~\text{K}$.
}
\end{figure}

\begin{table}
\caption{\label{tab:errors}The MAE, RMSE, the fitting error $\beta^{-1/2}$, and the average Bayesian error $\overline{\varepsilon}_{\text{B}}$ reported in units of [meV/\AA] for BAs, Si, and SnSe. The MAE, RMSE, and $\beta^{-1/2}$ have been evaluated on a test comprising 50 structures sampled from a $150~\text{ps}$ NVT trajectory at $300~\text{K}$, and the average Bayesian error $\overline{\varepsilon}_{\text{B}}$ on a $500~\text{ps}$ NVT trajectory.}
\begin{ruledtabular}
\begin{tabular}{lccc}
 & BAs & Si & SnSe\\
\hline
MAE & $7.0$ & 9.3 & 24.7 \\
RMSE & $8.8$ & 11.6 & 31.5 \\
$\beta^{-1/2}$ & $8.3$ & 10.1 & 30.0 \\
$\overline{\varepsilon}_{\text{B}}$ & 2.5 & 2.6 & 6.2\\
\end{tabular}
\end{ruledtabular}
\end{table}

In order to check the convergence of the dynamic training, we are going to follow the usual approach by evaluating the error of the predictions of the force fields with respect to \textit{ab initio} forces on a test set. For this purpose we have sampled 50 structures from $150~\text{ps}$ long NVT trajectories that were independent from the training trajectories. The following discussion will focus on BAs as an example. However, the same observation applies to Si and SnSe, for which the corresponding errors are listed in Table~\ref{tab:errors}. As can be seen in Figure~\ref{fig:figure_3} (a), the root mean square error (RMSE) as a function of the number of training structures quickly converges to a value of $8.8 ~ \text{meV/\AA}$ for BAs. While this is the case for both criteria considered in this study, we would like to note that the RMSE using our trajectory average criterion is always lower than that using the $\varepsilon_{\text{min}}$ criterion, validating our previous statement about the relative efficiency in improving the quality of the force field. It is noted that our RMSE of $8.8 ~ \text{meV/\AA}$ for the BAs force-field is considerably lower than that recently reported \cite{PhysRevMaterials.9.045403} using an MLIP trained with a static approach, which highlights the advantage of a dynamic training method. 

The RMSE of the forces on a test set described in the previous paragraph can be estimated by Equation~\ref{eq:sigma}. The second term in this equation, i.e. the Bayesian error $\varepsilon_{\text{B}}$, is positive semi-definite and will eventually go to zero for large numbers of training structures \cite{CBishop2006}. Thus, the RMSE is determined by the first term $\beta^{-1/2}$ representing the average error of the predictions of the force field with respect to the training data. Indeed, as shown in Figure~\ref{fig:figure_3} (b), $\beta^{-1/2}$ converges to its final value of $8.3 ~ \text{meV/\AA}$ as the number of training structures is increased, which is very close to the converged value of the RMSE. The remaining difference between $\beta^{-1/2}$ and the RMSE comes from the average of $\varepsilon_{\text{B}}$ over the test set. If that average of $\varepsilon_{\text{B}}$ is sufficiently small, adding additional training data will only marginally improve the RMSE with respect to \textit{ab initio} forces. Thus, in combination the fitting error $\beta^{-1/2}$ of the linear regression and the trajectory average of the Bayesian error $\mu_k[\varepsilon_{\text B}]$ evaluated on the training trajectory provide a cost efficient way to determine the state of convergence of the dynamic training. 

In the previous paragraph we have seen that the error of the forces is bound from below. However, that means that the force field will not reach arbitrary accuracy, even with very large amounts of training data. Thus, in order to further reduce the error of the force fields, we have to identify alternative ways to lower $\beta^{-1/2}$. First of all, the lower bound depends on the anharmonicity of the material and the temperature of the dynamic training. Both of these are related to the overall spread of the training data and cannot be changed in order to improve the force field. Besides these external dependencies, we expect the cluster expansion of the force field to play an important role, because in Bayesian linear regression the fitting error $\beta^{-1/2}$ represents contributions to the target value that are not captured by the basis functions of the regression model. Indeed, we find that removing the fourth order terms from the force field for BAs or reducing the cut-off radii for the third and fourth order terms for SnSe increased both $\beta^{-1/2}$ and the RMSE on the test set by $~\approx 50 \%$. We would like to note that in particular for BAs the increase of the RMSE by omitting the fourth order terms resulted in rare events during the MD simulations, similar to those described in \cite{PhysRevMaterials.9.063801}. In particular, in calculations for large supercells, we found erroneously predicted defect formation that led to the failure of the MD. On the other hand, adding even higher order terms or increasing the cut-off radii beyond the values reported in Table~\ref{tab:computational_details} did not significantly reduce the $\beta^{-1/2}$. This confirms that our force fields are already converged with respect to the cluster expansion.

\subsection{Distribution of uncertainties}

\begin{figure}[tb]
\includegraphics[width=\columnwidth]{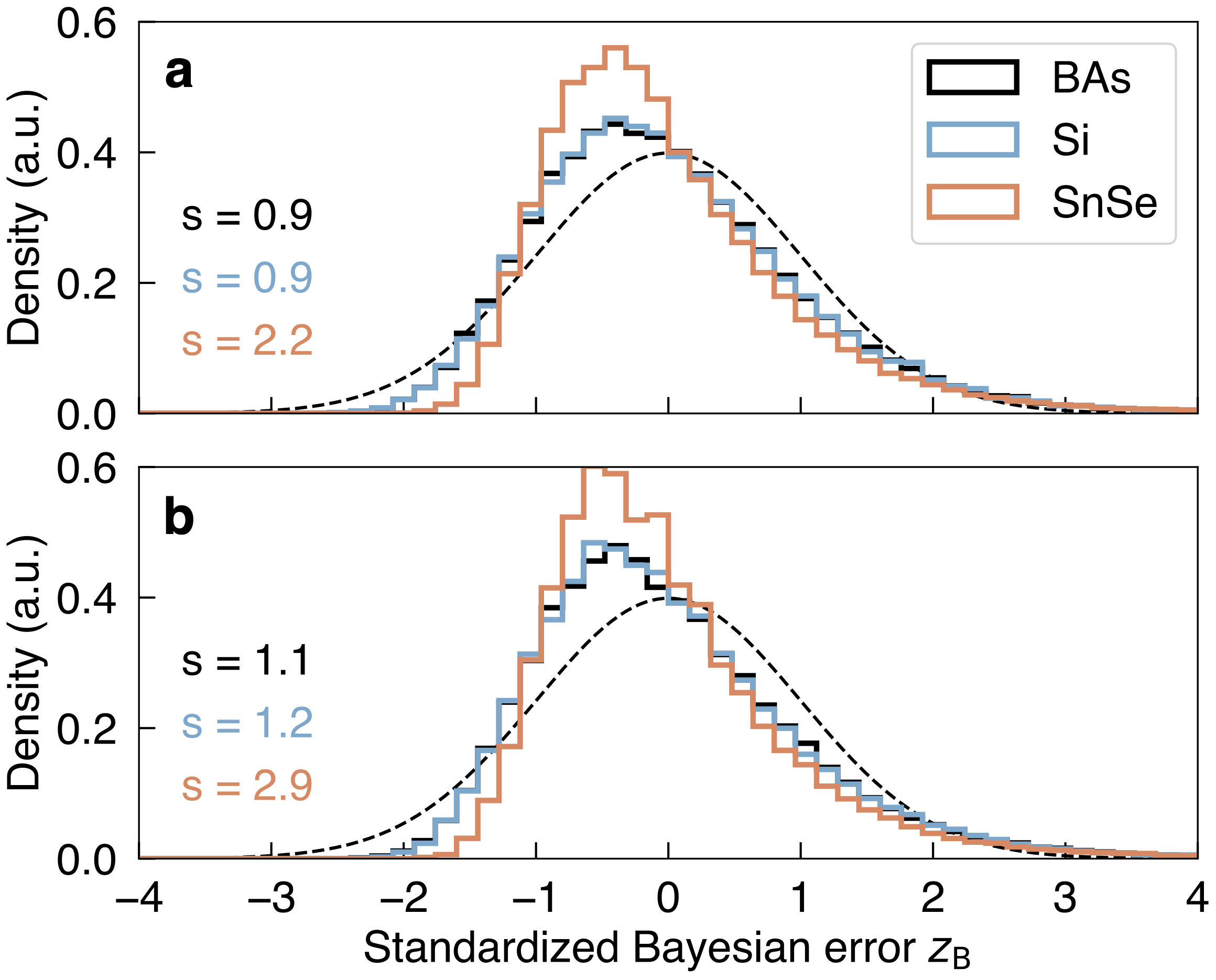}
\caption{
    \label{fig:figure_4}Distribution of the standardized Bayesian error $z_{\text B}$ over a $500~\text{ps}$ molecular dynamics trajectory for BAs (black), Si (blue), and SnSe (red).
     The employed force fields have been trained including either the full training data set (a) or a subset of the training data corresponding to a twice as large RMSE. An ideal normal distribution with mean $\mu = 0$ and standard deviation $\sigma = 1$ is shown for reference (dashed black). 
}
\end{figure}

As already mentioned earlier, our method to determine the convergence of the dynamic training requires that the uncertainty in the CS is sufficiently low to ensure that the fitting error of the linear regression is close enough to its converged value. So far we have relied on the average Bayesian error over a small test set or $\mu_k[\varepsilon_{\text B}]$ along the training trajectory to argue that the dynamic training of our force fields is converged. In order to confirm that this is not just a finite size effect due to a limited sampling of the CS, we are going to assess the overall distribution of uncertainties in the accessible part of the CS using MD. Obtaining a statistically meaningful sampling however, required relatively long, i.e. $500~\text{ps}$ NVT trajectories. The average Bayesian errors obtained from this more complete sampling of the CS are $2.5$, $2.6$, and $6.2 ~ \text{meV/\AA}$ for BAs, Si, and SnSe, respectively. This confirms that $\mu_k[\varepsilon_{\text B}]$ from the training trajectory and the average Bayesian error from the test set in the previous section are already good approximations for the overall uncertainty in the CS and thus that the dynamic training is sufficiently converged.

We can gain further insights by analyzing the distribution of uncertainties in the CS. For this purpose, we are going to standardize the Bayesian error according to $z_{\text B} = (\varepsilon_{\text B} - \mu)/\sigma$, where $\mu$ and $\sigma$ in this case are the mean and the standard deviation of $\varepsilon_{\text B}$ over the entire MD trajectory. This standardization leads to a distribution with zero mean and a standard deviation of $1$. As shown in Figure~\ref{fig:figure_4} the distributions of $z_{\text B}$ for BAs, Si, and SnSe are close to an ideal normal distribution, but slightly skewed towards the negative side of the axis. This is due to the added extra weight arising from the unusually high uncertainty in the less well represented regions of the CS. We can quantify to what extent the mean of the distribution is skewed away from the maximum representing the majority of structures along the MD trajectory by the skewness $s$, which is proportional to the third moment of the distribution. Even for SnSe, featuring the highest skewness of $2.2$, the uncertainty distribution is close enough to a normal distribution, which would be characterized by $s = 0$. In principle, the skewness of the uncertainty distribution can be an indicator for the convergence of the dynamic training as well. In particular, a small value of $s$ implies that there are not unusually many structures with $z_{\text{B}} > 3$ left for the dynamic training to find. This could be a useful alternative for methods, where the fitting error is expensive to evaluate, does not converge as smoothly as the $\beta^{-1/2}$ in the case of linear regression, or is only an average quantity that is not a good indicator for any specific element as in the case for universal MLIPs. However, we found that the skewness is not very sensitive with respect to the number of training structures. For BAs and Si training the force field with a subset of the full training data that corresponds to a twice as large RMSE still resulted in almost normally distributed uncertainty.

\section{\label{sec:summary} Conclusions}

In this work, we have showcased our dynamic training method that relies on Bayesian error estimation for efficiently selecting new training data for a machine-learning-based force field and simultaneously measuring the convergence of the training process. The decision about when to retrain the force field was informed by comparing the current Bayesian error to its trajectory average, which allowed us to identify new training structures with unusually high uncertainty within their respective regions of the CS. 

We have applied our dynamic training method with the trajectory average criterion to train force fields for materials with varying degrees of anharmonicity, in particular BAs, Si, and SnSe. We found that for all three materials under consideration the dynamic training method using the trajectory average criterion efficiently reduced the Bayesian error along the training trajectory as well as the RMSE with respect to \textit{ab intio} forces. Comparing the predictions of the force fields to \textit{ab intio} forces further revealed that the RMSE is bound from below by the fitting error of the linear regression, which converges towards the same final value as the number training structures is increased. In combination, the average Bayesian error and the fitting error define an interval within which the error of the forces is converged, thus providing an efficient way to measure the state of convergence of the dynamic training without requiring additional \textit{ab intio} calculations. 

By analyzing the overall distribution of uncertainty in the accessible part of the CS we confirmed that the fitting error of the Bayesian linear regression is already close to its converged value and thus that the dynamic training is sufficiently converged. In addition, the distribution of uncertainty itself provides information about the convergence of the dynamic training, since it evolves towards an ideal normal distribution as more and more training data are added. This provides an alternative way to assess the convergence of the dynamic training that only relies on the uncertainty measure. 

\section*{Acknowledgements}

We acknowledge support from Academia Sinica.

\bibliography{literature}

\end{document}